\documentclass[final,5p,times,twocolumn]{elsarticle}
\usepackage{amssymb}
\usepackage{booktabs}
\usepackage{tabularray}
\usepackage[table,xcdraw]{xcolor}

\newcommand{\sstar}[1]{$^{**}$}
\newcommand{\pb}[1]{$\dag$}
\newcommand{\ts}[1]{$\times$}
\usepackage{booktabs}

\begin{document}

\begin{frontmatter}

\title{Post-COVID Highlights: Challenges and Solutions of AI Techniques for Swift Identification of COVID-19}
\author[label1]{Yingying Fang}
\author[label2]{Xiaodan Xing}
\author[label1]{Shiyi Wang}
\author[label1]{Simon Walsh}
\author[label1,label2,label3,label4,label5]{Guang Yang}

\affiliation[label1]{organization={National Heart and Lung Institute, Imperial College London},
city={London},
postcode={SW7 2AZ},
country={UK}}

\affiliation[label2]{organization={Bioengineering Department, Imperial College London},
city={London},
postcode={W12 7SL},
country={UK}}

\affiliation[label3]{organization={Imperial-X, Imperial College London},
city={London},
postcode={W12 7SL},
country={UK}}

\affiliation[label4]{organization={Cardiovascular Research Centre, Royal Brompton Hospital},
city={London},
postcode={SW3 6NP},
country={UK}}

\affiliation[label5]{organization={School of Biomedical Engineering \& Imaging Sciences, King's College London},
city={London},
postcode={WC2R 2LS},
country={UK}}

\begin{abstract}

Since the onset of the COVID-19 pandemic in 2019, there has been a concerted effort to develop cost-effective, non-invasive, and rapid AI-based tools. These tools were intended to alleviate the burden on healthcare systems, control the rapid spread of the virus, and enhance intervention outcomes, all in response to this unprecedented global crisis. As we transition into a post-COVID era, we retrospectively evaluate these proposed studies and offer a review of the techniques employed in AI diagnostic models, with a focus on the solutions proposed for different challenges. This review endeavors to provide insights into the diverse solutions designed to address the multifaceted challenges that arose during the pandemic. By doing so, we aim to prepare the AI community for the development of AI tools tailored to address public health emergencies effectively.

\end{abstract}

\begin{keyword}
COVID-19  \sep swift identification  \sep AI-based diagnosis \sep 3D image processing \sep reliability \sep small datasets 
\end{keyword}

\end{frontmatter}

\section{Introduction}
\label{sec:intro}

By August 2023, COVID-19 had already claimed hundreds of millions of lives worldwide \cite{whodata}. At the onset of the pandemic, there was an urgent need for rapid and accurate detection to control the disease's rapid transmission, and AI emerged as a potentially faster, more sensitive, and readily available testing tool.

\begin{figure*}[ht]
\begin{center}
\includegraphics[width=\linewidth]
{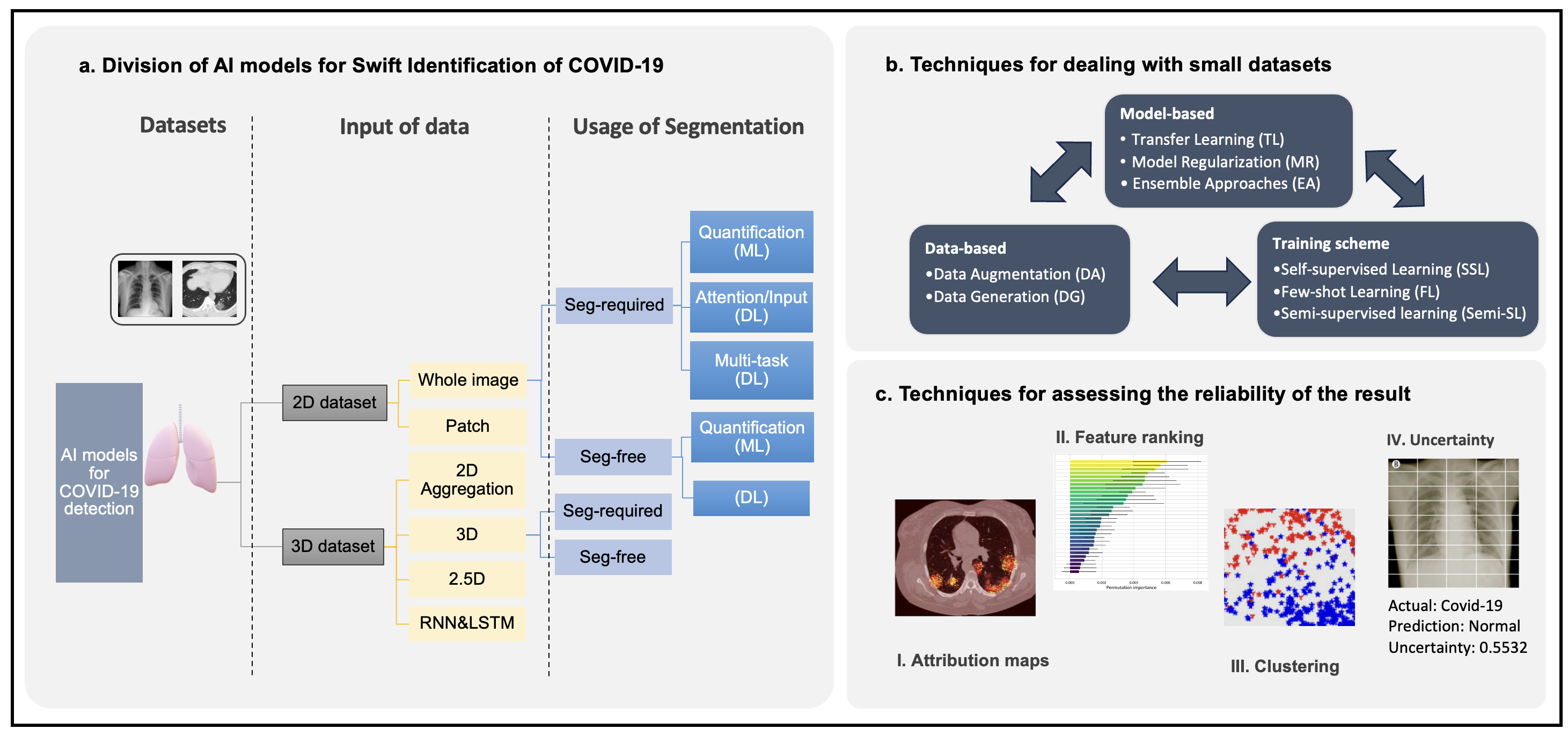}
\end{center}
\caption{\textbf{Overview of AI techniques in diagnostic models for COVID-19.} 
a) Model categorisation based on input format and segmentation necessity.
b) Enhancing model performance on small datasets through generalization techniques.
c) Methods for identifying predictive features and assessing prediction confidence to enhance result reliability.
}
\label{fig: overview}
\end{figure*}

However, in comparison to the remarkable strides AI has
made in healthcare sectors, the development of AI-based tools
for diagnosing or predicting outcomes in COVID-19 patients presents a unique and demanding set of challenges. Notably, the urgency of the pandemic requires an accelerated model development process, which demands the creation of AI models despite the limitations of scant data and an incomplete understanding of the disease. Meanwhile, the intricate pandemic environment where healthcare professionals confront an ever-evolving and poorly understood novel disease, also adds difficulty to the task of data labelling, increasing the complexity of
model training.

In this article, we provide an in-depth examination of various AI techniques that have emerged for the rapid diagnosis of this novel disease in this public health emergency. Compared to other reviews \cite{adadi_artificial_2022,roberts2021common}, our objective is to compare and analyse the effectiveness of these techniques in addressing diverse challenges, with the aim of assisting future researchers in selecting appropriate methods for specific situations and improving upon the limitations identified in current studies. This article is structured as follows: First, we categorise all methods according to the data dimensions and discuss the techniques used to transition from 2D CT slices to 3D CT volumes. Second, we categorise the methods into classes that require lesion segmentation and those that do not, emphasising the unique benefits of performing lesion segmentation as an auxiliary task. After categorising the existing methods, we delve into the strategies employed by these studies to tackle challenges related to small datasets and enhance the reliability of the developed models, which are evidently two significant challenges. Finally, we present a summary of limitations and future prospects within the reviewed studies. An overview of the key themes in this review is provided in Figure \ref{fig: overview} and a list of representative works is illustrated in Table \ref{tab:my_label}.

\section{Transition from 2D  to 3D  Analysis in COVID-19 CT Imaging}
In this section, we focus our attention on the works using 3D volumetric CT images, which provide the most detailed textures for monitoring the patient’s condition. Based on the manner in which the whole volumetric data were input into the model, we categorised the methods as Aggregation (3D-AG), whole-scan methods (3D-whole scan), and 2.5D methods (3D-2.5D).

\begin{itemize}
\item \textbf{3D-AG:} This approach refers to the methods which analyse 3D by aggregating the results from all 2D slices. Common aggregation ways include majority-decision which relies on a pretrained 2D model \cite{miron_covid_2021}, pooling operations which generate a global feature on the fully connected layer all the 2D slice \cite{zhang_transformer-based_2021,wang2020weakly,mei_artificial_2020,wu_deep_2020} or clustering methods \cite{qi_dr-mil_2021}. The major merit of such 2D-AG methods is their capability to leverage established and advanced 2D methods \cite{zhang_transformer-based_2021} which are limited by input size. However, this method is inherently limited to capturing 3D features between slices and fails to capture the complete 3D features within the scan.
\item \textbf{3D-Whole Scan:} On the other extreme, these methods directly input the entire scan into a 3D convolutional model after resizing the scans to a fixed number of slices, typically more than 64 \cite{harmon_artificial_2020,he2020benchmarking,hou_cmc-cov19d_2021}. This enables a comprehensive feature exploration across the complete scan. Nevertheless, this approach can be limited by memory constraints and is susceptible to overfitting, mainly because 3D convolutional models often involve a large number of parameters, while the available number of complete 3D scans is relatively small.
\item \textbf{3D-2.5D:} The 2.5D algorithms process a limited number of slices, typically ranging from 10 to 48, which are significantly downsampled from the entire scan, and input into the network as a unified entity \cite{he2020benchmarking,juarez_covid-19_2023,harmon_artificial_2020}.
This approach aims to harness the benefits of a more comprehensive feature analysis within 3D methods while concurrently reducing the computational memory required by such methods.
\end{itemize}

\begin{table*}
\caption{\textbf{An overview of the representative methods discussed in the article.} Please note that unless specified, dataset size is illustrated by patients. Open datasets are identified by their names without patient counts. In cases of small datasets, multiple methods may be employed, but this table highlights major techniques applied in each paper (not an exhaustive list). $^*$ denotes representative works and  $^{**}$ denotes works of outstanding interest.}
\label{tab:my_label}
\small
\centering
\resizebox{0.75\textwidth}{!}
{
\begin{tblr}{llllll}
\hline
\textbf{Literature} & \textbf{Datasets} & \textbf{Input} &  \textbf{Segmentation} &  \textbf{Small Dataset Technique}& \textbf{XAI}  \\ 
\hline
\SetCell[c=6]{l}{\textbf{3D inputs: 3D-AG, 3D-Whole Scan, 3D-2.5G}}\\
\cite{zhang_transformer-based_2021} & COV19-CT-DB & {CT: 3D-AG} & / & / &/ \\
\cite{miron_covid_2021}  & COV19-CT-DB & {CT: 3D-AG\&Whole Scan}&/& {TL,EA\&DA} & / \\
\cite{hou_cmc-cov19d_2021}** & COV19-CT-DB & {CT: 3D-Whole Scan}&/& \textbf{TL,EA\&DA,MixUP\&SSL} & Grad-CAM \\
\cite{he2020benchmarking}**  & {CLEAN-CC-CCII} & {CT: 3D-Whole Scan\&2.5D}& / & TL\&\textbf{MixUP} &  CAM  \\
\cite{mei_artificial_2020}* &  {Covid: 419 \\Non-Covid: 486} & CT: 3D-AG & / & / & Attribution map\\
\cite{wu_deep_2020} &  {(Patient) \\ Covid: 368 \\ Pneumonia: 127} & CT: 3D-AG & / & / & /\\
\cite{juarez_covid-19_2023} & {CC-CCII dataset}  &{CT:3D-2.5D\\} & / & {TL}&  /  \\
\cite{harmon_artificial_2020}*   & {Covid: 922 \\ Non-Covid: 1625} & {CT: 3D-Whole Scan\&2.5D}& / &  {DA, Resampling} & Grad-CAM  \\
\cite{meng2023bilateral}** & {{Cleaned CC-CCII\\ COVID-CTset\\MosMed}} & CT: 3D-2.5D & / & / & {\textbf{Uncertainty}\\ Grad-CAM\\Graph reasoning} \\
\cite{qi_dr-mil_2021}** & {Covid: 141\\ CAP: 100} & CT: 3D-AG & / & model regularisation \& DA & Clustering \& Grad-CAM \\
\hline
\SetCell[c=6]{l}{\textbf{Segmentation: Quantification, Attention, Multi-task}}\\
\cite{fang_ct_2020}  &  {COVID: 46 \\ Pneumonia: 29}  & {CT: 2D} &  {Radiomics} & Dimension reduction & {Clustering \& Feature ranking}   \\
\cite{liu_ct_2021}*  &  {COVID: 115\\ Viral pneumonia: 435}  & {CT: 3D-mean result of multi-lesions} &  {Radiomics} & {Dimension reduction} & {Feature ranking} \\
\cite{chen_machine_2021}** &{COVID: 63 \\Pneumonia: 71 }& {CT:3D-merged ROI}  & {Radiomics\\ other quantification\\} & {Dimension reduction}& {Feature ranking}\\
\cite{zhang2020clinically}* & {Covid: 752 \\ Pneumonia: 797 \\  Normal: 697 } & {CT: 3D-Whole Scan} & {Input + Quantification \\ } & / & {Attribution\&Feature ranking} \\
\cite{wang2020weakly}*  & {Covid: 313\\ Non-covid: 226} & {CT: 3D-AG\&3D} & {Input\\} & DA & CAM; 3DCC  \\
\cite{gao_dual-branch_2021} & {(Patient) \\ Covid: 704  \\ Normal: 498 \\Total slices: 210,395} & {CT: 3D-AG} & {Attention \& Multi-task \\} & / & {Segmentation} \\
\cite{zeng_ss-tbn_2023}** & {Covid: 48  \\ Normal: 75\\}  & {CT: 3D-AG\\} & {Attention \& Multi-task} & \textbf{Semi-SL} &{Segmentation}  \\
\cite{wang_joint_2021}**  & {3DLSC-COVID} & {CT: 3D-2.5D} & {{Input\&Multi-task \\ \& Quantification}} & / &  {\textbf{Segmentation, CAM}\\  \textbf{Quantification}}\\
\cite{wu_jcs_2021} & {Covid: 400 \\  Normal: 350 \\ Total slices: 144,167} & {CT: 2D} & {Seg-required: Multi-task} & / & {Grad-CAM, Segmentation} \\
\cite{hu_weakly_2020} & {Covid: 150 \\  Normal: 150 \\ CAP: 150} & {CT: 2D\\}  & {Multi-task} & \textbf{Weakly-SL}  & {Integrated Gradient}   \\
\cite{li_explainable_2022}& { Covid: 140 \\CAP: 124 \\ Normal: 115}& CT: 2D & {Multi-task\\}& \textbf{Semi-SL}& {CAAM,CAM,Segmentation} &\\
\hline
\SetCell[c=6]{l}{\textbf{Small Dataset Techniques: Model-based, Data-based, Training-based}}\\
\cite{he_sample-efficient_2020}** & {Covid: 349 slices \\ Non-Covid: 397 slices} & {CT: 2D}& /&{TL, SSL} & {Grad-CAM} \\
\cite{wang_covid-19_2021} & {Covid:320\\ Normal: 320} & {CT: 2.5D\\}& / & {TL\&DA} &{ Grad-CAM} \\
\cite{loey_within_2020} & {Covid: 69 CXR\\ Normal: 79 CXR\\  Pneumonia bacterial: 79 CXR\\ Pneumonia virus: 79 CXR} & {CXR: 2D}& / & {TL \& Synthesis} & / \\
\cite{farkas_covit-gan_2021}** & {COVID-CT\\Sars-CoV-2} & {CT: 2D, X-Ray: 2D?}& / & {TF; Synthesis} & {Attention Map}  \\
\cite{jiang_few-shot_2021}** & {COVID-19 CT} & {CT: 2D}& / & {Synthesis} & / \\
\cite{zhang2023gionet}** & {COVIDx CT-3} & {CT: 3D\\}& / & {TL \& Synthesis} & {\textbf{Clustering}, Activation Map}  \\
\cite{chen_momentum_2021} & {COVID: 216 slices\\ Non-Covid: 274 slices} & CT: 2D & / & FL & Grad-CAM \\
\cite{ornob_covidexpert_2023} & {Covid: 200 slices \\  Normal: 200 slices\\ CAP: 200 slices } & CT: 2D & / & EA\&FL& / \\
\cite{haque_generalized_2021} & {Covid: 1583 CXR\\ Normal: 4273 CXR}  & Xray: 2D & / & {DA,MixUP\&Semi-SL} & Saliency maps \\
\cite{aviles2022graphxcovid}* & {15,254 CXR: \\ Covid; Pneumonia; Normal}   &Xray: 2D & / &{Semi-SL}&\textbf{Attribution map\&Uncertainty}\\
\cite{zeng_ss-tbn_2023}* & {Covid: 48  \\ Normal: 75\\} & {CT: 3D-AG\\} & {Attention \& Multi-task} & \textbf{Semi-SL} &{Segmentation}  \\
\hline
\SetCell[c=6]{l}{\textbf{Reliability: Attribution maps,Feature ranking, CLustering, Quantification Measurement}}\\
\cite{minaee_deep-covid_2020} & COVID-Xray-5k & Xray: 2D & / & TL & Attribution maps \\
\cite{gong2022explainable} & {Covid: 759\\ Non-Covid:978} & clinical data  & / & Ensemble & Feature ranking \\
\cite{shamsi2021uncertainty}* & {(Xray+CT) \\ Covid: 25 Xray + 349 slices \\ Non-covid: 73 Xray + slices CT} &  {Xray\&CT: 2D} & / & TL & Uncertainty, Grad-CAM \\
\cite{meng_bilateral_2023}** & {CC-CCII\\COVID-CTset} & CT: 3D-AG& Input & / & Uncertainty,  Grad-CAM \\
\hline
\end{tblr}}
\end{table*}

Although each has its own advantages, comprehensive and equitable comparisons of these input methods have been scarce in existing literature \cite{qi_dr-mil_2021}. Some studies have demonstrated superior performance and lower computation cost for 2.5D and whole-scan methods compared to AG  \cite{wang2020weakly} and some have reported comparable performance of whole-scan and 2.5D input with multiple sampling \cite{harmon_artificial_2020, he2020benchmarking} whereas contrary conclusion is found in \cite{qi_dr-mil_2021}.
It is worth mentioning that assessing the efficacy of diverse input across varying training dataset sizes is imperative. For example, 3D-Whole Scan methods can be susceptible to overfitting with smaller datasets, while they perform better in larger datasets.
\section{How does segmentation help for the diagnostic models?}
Segmentation is typically used to delineate regions of interest (ROIs) such as tumours or lesions in medical images. Depending on whether 2D and 3D methods incorporate lesion segmentation in classification models, we categorise them as either `seg-free' or `seg-required' methods. Achieving accurate segmentation often requires a substantial investment of time in ROI annotations. Hence, we evaluate the potential benefits and methods of integrating segmentation into AI-based diagnostic models. This assessment provides insights for algorithm developers deciding whether to include this step in their circumstances.

\begin{itemize}
\item \textbf{Quantification of Lesion Features:} Extracting lesion-related features is the primary purpose of segmentation by quantifying the segmented area into an array of features. These features, either analysed by radiologists manually \cite{fang_radiomics_2020,chen_machine_2021,wang_joint_2021} or computer automatically \cite{fang_ct_2020,fang_radiomics_2020,liu_ct_2021,chen_machine_2021,zhang_deep_2021}, are then utilised by machine learning models for the classification task. The major advantage of this method is that these quantified features input into the model are human-interpretable compared to deep learning models, which take the entire image as input without specifying which specific features are being utilised \cite{mei_artificial_2020}.
\item \textbf{Attention constraints:} Utilising segmented area as attention area in deep learning models is another strategy to leverage segmentation results \cite{zhang2020clinically,wang2020weakly,gao_dual-branch_2021,zeng_ss-tbn_2023}. This approach enables the model to focus on regions deemed significant based on human knowledge, ultimately enhancing the reliability of predictions by eliminating bias that may exist in unrelated areas.
\item \textbf{Multi-task model:} A third approach entails training a multi-task model capable of simultaneously performing lesion segmentation to localise the affected area and patient classification \cite{gao_dual-branch_2021,amyar2020multi,wang_joint_2021,wu_jcs_2021,hu_weakly_2020,li_explainable_2022}. This design enhances the information interaction between the two related tasks, which aims to bring performance gain to both tasks \cite{gao_dual-branch_2021,wang_joint_2021}. Moreover, the segmentation result can serve as support to explain the classification results.
\end{itemize}
To conclude, the primary advantage of incorporating segmentation into the COVID-19 identification workflow is to enhance the explainability and reliability of AI models. Various methods were explored to achieve segmentation during the pandemic. These approaches range from manual delineation \cite{fang_ct_2020,liu_ct_2021} and the utilisation of off-the-shelf segmentation algorithms \cite{fang_radiomics_2020,chen_machine_2021} to the customization of data-driven deep learning-based methods with designed ROIs \cite{zhang_deep_2021,zhang2020clinically,gao_dual-branch_2021,zeng_ss-tbn_2023}. Additionally, efforts were made to develop methods that alleviate the necessity for manual labeling for lesion localization.

\section{How to dance with small data?}

At the beginning of the pandemic, limited data for model training posed a risk of overfitting. While data scarcity is not a new challenge in the medical field, the pandemic has brought various approaches together to address this issue. We categorise these different methods into model-based, data-based, and learning-based schemes.

\begin{itemize}

\item \textbf{Model-based scheme:}  We categorise model configuration methods as model-based methods. Among them, Transfer learning (TF) stands out as a widely adopted method. It utilises a pretrained model parameter from a larger source dataset as initialized weights for training a target model on a smaller dataset, consistently improving model performance in many studies \cite{he_sample-efficient_2020, loey_within_2020, hou_cmc-cov19d_2021}. Additionally, other methods also include ensemble approaches (EA), where multiple models are combined to enhance the generalisation and reliability of the model’s predictions \cite{maftouni2021robust}, and regularisation techniques applied to the model’s parameters to prevent overfitting.
\item \textbf{Data-based scheme:} This involves data manipulation methods to boost training data volume, mainly through techniques such as data augmentation (DA) and data generation (DG). DA includes basic transformations applied to the original data to increase the number of training data, while it cannot change the underlying statistical data distribution \cite{wang_covid-19_2021, wang_psspnn_2021}. In contrast, DG aims to generate new data within the training dataset, potentially improving the overall data distribution. Conventional data generation techniques include MixUP \cite{he2020benchmarking, hou_cmc-cov19d_2021}, which creates new data by combining random image pairs. More recent AI techniques involve synthesising a wealth of high-quality images using deep generative models \cite{jiang2020covid, xing2023less, xing2022cs}. Several studies have investigated the performance gain in adopting these synthesised images into their datasets for COVID-19 identification tasks \cite{loey_within_2020, farkas_covit-gan_2021, jiang_few-shot_2021, zhang2023gionet}.
\item \textbf{Learning-based scheme:} Learning-based method is a group of methods that are performed during the training of a deep learning model. Self-supervised learning (SSL) focuses on enhancing a model’s parameterisation by training it to learn generalised representations through pretext tasks without using task labels \cite{chen_momentum_2021}. Few-shot learning (FL), a training technique specifically designed to tackle the problem of insufficient training data has been applied and validated in COVID-19 datasets \cite{chen_momentum_2021, shorfuzzaman2021metacovid, jiang2021few, jadon_covid-19_2021, wang2022semantic, ornob2023covidexpert}. Another solution for addressing small datasets is semi-supervised learning (Semi-SL) \cite{zeng_ss-tbn_2023,li_explainable_2022,aviles2022graphxcovid, haque_generalized_2021}, which leverages the unlabelled data in the training procedure.

\end{itemize}

It’s noteworthy that while most DL models can utilise various methods within or from different classes to maximize performance, current machine learning models relying on quantified features resort only to reducing feature dimension or ensembling models to avoid overfitting.

\section{Can we can trust the model?}

Evaluating the reliability of AI models remains an ongoing question in medical applications. In this review, we have assessed the methods used to enhance the reliability of AI-based COVID-19 models and categorized them into four distinct groups.

\begin{itemize}
\item \textbf{Attribution map-based methods:} These methods aim to return an attribution map for inout images,  which highlight the the relevant regions for the prediction. It can be achived through various ways including the class activation maps (CAM) \cite{harmon_artificial_2020,hou_cmc-cov19d_2021,wang2020weakly,wu_jcs_2021,li_explainable_2022,he2020benchmarking}, gradient-based methods  \cite{hu_weakly_2020,haque_generalized_2021}, perturbation methods \cite{minaee_deep-covid_2020} and segmentation results \cite{wu_jcs_2021,li_explainable_2022}.  Attribution maps are used to verify if the models utilise relevant areas rather than biased features.
\item \textbf{Feature ranking:}  In machine learning models, finding the most important features for the decisions can also help interprete the model decisions. To achieve this goal, methods such as statistical analysis  \cite{yu2020multicenter}, LIME \cite{gong2022explainable}, SHAP \cite{zhang2020clinically} and model weights are used to rank the features.
\item \textbf{Clustering methods:} The clustering methods explain the model prediction by providing similar examples as reference to the test cases  \cite{fang_ct_2020,yamga2022identifying,qi_dr-mil_2021}. This technique is promising to enable the users to discover common patterns from the near samples.
\item \textbf{Uncertainty Measurements:} Instead of explaining the model, recent studies have introduced various methods to assess the model’s confidence in its predictions \cite{ghoshal2020estimating,khosravi2022uncertainty,shamsi2021uncertainty,meng_bilateral_2023,meng2023bilateral}. This allows model users to identify uncertain cases for further examination by radiologists and hence increase the reliability of the predictions with high confidence. By applying this technique, the models provides more reliable predictions.
\end{itemize}

Among these methods, attribution maps and feature mapping aim to elucidate model decisions by identifying crucial features. While attribution maps effectively detect lesions in infectious cases, they struggle to differentiate COVID predictions from other pneumonia types, as both classes exhibit similar lesion patterns. Furthermore, different calculation methods for these explanations can yield significantly divergent results within the same model, raising concerns about the reliability of attribution maps themselves \cite{ghoshal2020estimating}. These issues limit the effectiveness of these explanation methods in assessing model reliability. Clustering explanation, successful in 2D image analysis \cite{wulczyn2021interpretable}, remains relatively unexplored in COVID-19 and may face added challenges in the 3D volume context. In light of these challenges, uncertainty estimation presents a promising solution applicable to all models, enhancing overall reliability.

\section{Limitation and future studies}

In this section, we conclude several limitations identified in the literature we have reviewed and offer insights into potential areas for future research:

\begin{itemize}
\item Despite COVID-19's novelty, its lung manifestations can resemble other conditions. Utilising off-the-shelf segmentation methods to detect distinct lung abnormalities could expedite the development of required segmentation methods.
\item Variations in lesion selection, segmentation, feature extraction, and machine learning models have impeded consensus on critical features for COVID-19 identification in presented radiomics-based methods. This challenge is exacerbated by the selection of lesions characterized by scattered patterns across scans and the use of extremely small datasets. To address this, further validation on larger datasets and exploration of additional AI-based quantification methods are essential.
\item AI holds the potential for discovering new biomarkers \cite{wang_joint_2021,yu2020multicenter}, but more reliable attribution maps and feature ranking methods are needed to support advanced AI models in this pursuit. The reliability of attribution methods remains an understudied area in the realm of medical image analysis.
\item  The effectiveness of advanced AI techniques, such as few-shot learning and data synthesis, requires further validation on larger independent datasets and additional development in the context of 3D input methods.

\end{itemize}

\section{Conclusion}

This work reviews AI-based models that have emerged for the rapid detection of COVID-19 by examining solutions related to the development of such tools. We enumerate and compare the various methods that have been deployed. Our analysis of these diverse techniques, including their strengths and limitations, aims to benefit the AI community by enhancing pandemic preparedness. Furthermore, this review also provides valuable insights for AI diagnostic tools applied to lung diseases, such as ILD, which often present scattered lesions in lung scans.

\section*{Acknowledgement}
This study was supported in part by the ERC IMI (101005122), the H2020 (952172), the MRC (MC/PC/21013), the Royal Society (IEC\textbackslash NSFC\textbackslash 211235), the NVIDIA Academic Hardware Grant Program, the SABER project supported by Boehringer Ingelheim Ltd, Wellcome Leap Dynamic Resilience, and the UKRI Future Leaders Fellowship (MR/V023799/1).

\bibliographystyle{elsarticle-num}
\bibliography{revised}
\end{document}